# Sustainable Development Goal Target Interactions in the Philippines: A Two-Method Approach

Bongolan, V., Soria, S., Rivera, R.

*System Modelling and Simulation Laboratory*
*Department of Computer Science*
*University of the Philippines—Diliman*

## Abstract

In 2015, the United Nations adopted 17 Sustainable Development Goals (SDGs) with 169 targets for transformation toward a more sustainable future by 2030. This study seeks to evaluate and analyze SDG target interactions in the Philippines to determine resolution measures for conflicting targets, and prioritization for targets that reinforce others and have no conflicts. To evaluate all 14196 target interactions, two methods are employed. First, experts with over five years of SDG-related experience evaluated interactions using a 7-point scale. Second, a non-parametric Spearman rank correlation is used on official indicator data with resulting coefficients serving as interaction scores. Interaction scores are then interpreted to mean that target pairs interact positively (synergies), negatively (trade-offs), or neutrally (nonclassified). Targets are also modelled as nodes and interactions as edges in graphs presented in sdg-interactions.herokuapp.com.

Results from the two methods were synthesized to formulate recommendations for concerned parties. This includes resolving negative intra-goal target interactions involving targets 3.1 "Reduce maternal mortality", 3.6 "Reduce road injuries and deaths", and 3.7 "Universal access to sexual and reproductive care, family planning, and education". Ugly targets (at least one negative interaction) including target 3.6, 3.7, and 8.2 "Diversify, innovate, and upgrade for economic productivity" also need to be resolved.

Targets that reinforce their corresponding SDGs should be prioritized, including 1.1 "Eradicate extreme poverty", 1.2 "Reduce poverty by at least 50%", 4.2 "Equal access to quality pre-primary education", 4.B "Expand higher education scholarships for developing countries", 6.2 "End open defecation and provide access to sanitation and hygiene", 6.6 "Protect and restore water-related ecosystems", 8.1 "Sustainable economic growth", 8.5 "Full employment and decent work with equal pay", 9.4 "Upgrade all industries and infrastructures for sustainability", and 9.5 "Enhance research and upgrade industrial technologies". Beautiful targets (no negative interactions) should also be prioritized, including target 8.5 and 17.5 "Invest in least developed countries".

## I. Introduction

In 2015, the countries of the United Nations agreed to adopt the 2030 Agenda for Sustainable Development in the United Nations General Assembly. This agenda, often referred to as Agenda 2030, endeavors to be a framework through which a better future can be secured for the global community. This is to be accomplished by addressing issues of health, education, and environmental preservation, all in a concerted effort towards sustainability. These global issues will be addressed by focusing on the 17 Sustainable Development Goals, commonly referred to as SDGs, with 169 specific targets that all involved countries should strive to realize by the year 2030.

At present, the Philippines is still considered to be a third world country. Plagued with corruption in the government, a lack of quality education, and poor access to healthcare, the country continues to struggle in its overall development. The 17 Sustainable Development Goals of Agenda 2030 were designed by the United Nations to be a model through which all countries regardless of development status can work towards a better and more sustainable future. The goal of this study is to identify what priorities should be made to achieve this agenda. To this end, the study seeks to identify which targets cause conflicts or trade-offs with other targets and which targets positively reinforce others, both within the same goal (intra-goal) or across different goals (inter-goal).

In order to do this, the study approaches the problem via two methods, viz: a) by a panel of experts (ISC, 2017) and b) analyzing official indicator data (Pradhan et al., 2017). The respondents for the first method will provide an evaluation on the inter- and intra-goal interactions between the 169 SDG targets. They are also to provide professional insight regarding negative interactions in order to guide policymakers on how these conflicts can be resolved for the country to carry out the 2030 Agenda.

An analysis of SDG target interactions can assist in directing the country's efforts to achieve the 2030 Agenda by taking advantage of positive target interactions as well as pointing out negative interactions that need to be further investigated or resolved. By limiting analysis to the Philippine context, insight on performance in terms of the 2030 Agenda would be better suited to the country's status.

[Research gaps to be added]

## II. Data Gathering and Analytical Methods

Previous studies have employed two main methods to evaluate SDG interactions namely: expert evaluation (ISC, 2017) and official indicator data (Pradhan et al., 2017). While previous studies have used these methods separately, this study uses both in order to draw a synthesized conclusion. This is primarily motivated by the reason that the study initially began using only the first method, expert evaluation. Since data gathering for this method was slow-going, there was an interest in exploring other methods to evaluate more target interactions towards a more complete analysis.

By having two different approaches to the same problem, the study aims to provide recommendations backed by two data sources regarding the implementation of positive interactions or the resolution of negative interactions. This is done by synthesizing results from both methods, identifying similar results to put forward a positive answer highlighting targets to prioritize and a negative answer highlighting targets that need to be resolved.

**2.1. Data Gathering and Interpretation**

2.1.1. First Method: Expert Evaluation

For the first method, the use of expert evaluation, a survey was conducted among experts from various fields relating to the SDGs. A discussion on how the web application for the survey was designed and implemented can be found in Section III.

To ensure the credibility of the SDG target interaction scores, respondents are screened by the study's curators. This is done initially by an invitation from the curators of the study to an interview to determine whether potential respondents are qualified to make target interaction evaluations. These potential respondents are then asked to create an account via the sign-up page of the web application where they indicate if they have five or more years of experience in their respective fields related to the advancement of the SDGs. A view of the sign-up page with all the necessary information required can be seen in Appendix: Figure 1.

Upon application, each respondent is required to indicate who among the curators invited them. This curator will then be notified via email and on their administrator dashboard of a pending account approval request. Once approved, a respondent will then be able to log into the website where they will be asked to choose a minimum of two SDGs that are best aligned with their expertise. A view of SDG expertise selection can be found in the Appendix: Figure 2. From their selected SDGs, the system will then generate a series of SDG target pairs for them to evaluate.

Each SDG target pair generated by the system represents an interaction between two SDG targets, and the users will then have to score each pair positive or negative using the same

7-point scale used by the International Science Council (2017). The seven points on the scale are as follows: -3 for cancelling, -2 for counteracting, -1 for constraining, 0 for consistent, +1 for enabling, +2 for reinforcing, and +3 for indivisible. A view of how an SDG target pair will be scored can be seen in the Appendix: Figure 3.

For each negative evaluation, the respondent is required to provide an explanation for their response. Explanations for positive evaluations are optional. These explanations are meant to provide insight into the nature of the target interactions as well as what factors may have caused them to merit such an interaction. For negative interactions, an explanation from the respondents may also provide ideas on how the negative interaction can be resolved. Users have the option to skip SDG target interactions that they do not wish to evaluate immediately to allow for flexibility. As long as they have not finalized their answers by submitting them, they can come back at a later time to provide their scoring.

SDG target pairs are bound to their user and will not be reassigned to a different respondent. Thus, all respondents will be required to answer all target pairs given to them, underscoring the importance of thoroughly screening and briefing respondents to provide a reliable evaluation.

Each SDG target pair is to be scored only once, which means that the score given by the respondents is final. Provisions are made to allow users to review and edit their answers. However, once they submit their answers, the target interaction scores will automatically be saved to the database and immediately reflected in the results pages of the web application, including all related graphs.

### 2.1.2. Second Method: Official UN Indicator Data

The second method makes use of official indicator data publicly provided by the United Nations Statistics Division. The indicator data available are based on a time-series from 1990 through 2018 and are not complete. Applying the method in Pradhan's study (2017), [Anderson et al., 2021] used the 2018 indicator data and passed it through a non-parametric Spearman rank correlation. The resulting correlation coefficients were what was used in this study to determine a score for target interactions. Scores were then interpreted with the same threshold as Pradhan and Anderson's studies. That is, for coefficients less than or equal to -0.6, the interaction is considered to be a trade-off. For coefficients greater than or equal to 0.6, the interaction is considered to be a synergy. Values in between -0.6 and 0.6 as well as those indicator pairs that do not have any available data were considered to be nonclassified. From there, the percentage of synergies, trade-offs, and non-classified indicator pair interactions at the target level was used to interpret whether a target pair was synergistic, a trade-off, or neither. That is, if a majority of the indicator pair scores are synergies then the target pair is synergistic, and so on.

**2.2. Analytical Methods**

The study looks into two things as part of its analysis of results across both data gathering methods: the nature of intra-goal target interactions and the ugliest and most beautiful targets, both explained hereafter. Results from both these analytical methods across both data gathering methods were then used to come up with a positive and negative answer.

2.2.1. Intra-Goal Target Interactions

As the name implies, intra-goal target interactions refer to those interactions between targets under the same SDG. The primary motivation for this is to determine whether the specific targets actually help in the advancement of their corresponding SDGs or whether they are in conflict with other targets of the same SDG. The implication is that negative intra-goal interactions should be resolved while positive ones should be reinforced.

2.2.2. Ugliest and Most Beautiful Targets

The study also looks into what we refer to as ugly and beautiful targets. We define an ugly target as an SDG target that has at least one negative interaction with another target, whether intra-goal or inter-goal. On the other hand, a beautiful target is an SDG target with no negative interactions. We determine a target's "ugliness" or "beautifulness" by the number of negative and positive interactions they have respectively. Thus, the more negative interactions a target has, the uglier it is, and the more positive interactions it has the more beautiful it is.

The motivation for this analysis is firstly to determine which targets need to be prioritized such that, based on available data, they are guaranteed not to have any conflicts with other targets. A list of most beautiful targets provide a priority list of targets to be reinforced. Another motivation for this analysis is to identify the ugliest targets that need to be resolved. Having only a few targets to focus on allows for a straightforward recommendation for parties involved in the SDGs.

2.2.3. Integration of Results for a Positive and Negative Answer

In order to come up with a positive answer, the study looks at positive interactions from expert evaluation and synergies from official indicator data. Similar target pairs among positive/synergistic intra-goal target interactions are then identified as well as the common targets in the lists of the most beautiful targets. These targets or target pairs will then be recommended for prioritization.

For the negative answer, the study looks at negative interactions from expert evaluation and trade-offs from official indicator data. Similar target pairs among negative/trade-off intra-goal target interactions are also identified as well as common targets in the lists of the ugliest targets. These targets or target pairs will then be recommended to be resolved.

## III. Web Application

In order to gather data for the first method and display the results for both methods, a web application was developed and deployed to Heroku at the address [http://sdg-interactions.herokuapp.com/](http://sdg-interactions.herokuapp.com/). This application is divided into two main components discussed hereafter:

1. Front-end (user interface and design): ReactJS, a JavaScript framework
2. Back-end (data management): API (Application Programming Interface) developed using Flask, a Python framework.

The front and back-end components were developed separately, both deployed to Heroku, with the front end serving as the home of the main website.

### 3.1. Web Application Development

3.1.1. Front-End Development: React

React was used for the front-end development of the website. React is a JavaScript library used for making user interfaces on mobile and web applications. It was used to design user and administrator views with their corresponding functionalities as shown on Table 1.

| ADMINISTRATOR | USER |
|---|---|
| Confirm Pending Users | Add Goals |
| View Dashboard | View Front Page |
| View Graph | Log In |

| View Menu Bar | View Menu Bar |
|---|---|
| View All Answers | Review Answers |
| View Users | Select Goals |
|  | View Settings |
|  | Sign Up |
|  | View Survey |
|  | View Tabs |

*Table 1. Admin and User privileges*

Reactstrap was also used to integrate needed components from Bootstrap. Axios, a Promise-based HTTP client widely used by React developers, was also utilized in order to connect to the database. The most significant tool used for the application was the react-d3-graph library, which makes possible the creation and design of the graphs needed to analyze the results of the study.

3.1.2. Back-end Development: Flask

The back-end of the system serves up a RESTful API that collects, saves, and processes the data and was developed on top of Flask. The API requests handle user creation and data gathering.

PostgreSQL is the database management system used, interfaced with the Python SQL toolkit, SQLAlchemy. Four models are used in the system: the User model, the SDG model, the User to SDG model, and the Survey Answers model.

All necessary information about the respondents are stored in the User model. Aside from login credentials, it also contains information on educational attainment, years of experience, organizational affiliations, and preferences on whether respondents would like to be acknowledged in the study.

SDG target descriptions are stored in the SDG model to serve as reference for the users while taking the survey or viewing the graphs and other result features. The User to SDG model links respondents to their chosen SDGs while the Survey Answers model stores generated SDG target interaction pairs and their corresponding evaluation scores given by the users. The APIs created handle the processes to be discussed in Data Gathering.

## 3.2. Website Features

### 3.2.1. Graph Views

The current progress of the study, based on how much data has been gathered so far, can be viewed by the public on the website. One does not need to have an account in order to view the graphs and other results of the study.

A graphical representation is used by the study in order to capture the nature of different target interactions based on the interpretation of results discussed in Section 2.1. The graphs portray SDG targets as nodes while the edges that connect them represent the target interactions. Nodes are color-coded based on the official colors of their corresponding SDGs used in the UN SDG website. That is, nodes for SDG 1 targets are colored red, nodes for SDG 7 targets are colored yellow, nodes for SDG 9 targets are colored orange, and so on. The color of each SDG is shown in Appendix: Figure 4. The nodes are also labeled with their SDG and target number. This allows users to easily distinguish targets apart.

For the results of expert evaluation which uses the 7-point scale, edges representing target interactions that have been evaluated to be positive (+1, +2, and +3) are colored blue, negative interactions (-1, -2, and -3) are red, and zero (0) interactions are black. Different shades of blue and red are used for the links corresponding to positive or negative interactions respectively. That is, the more positive or the more negative the interaction is, the darker the shade of the color (darker blue or darker red) used for the edge. Edges representing unevaluated target interactions are colored gray.

For the results of official indicator data, links for synergies (greater than or equal to 0.6), trade-offs (less than or equal to -0.6), and nonclassified (between -0.6 and 0.6) interactions are simply colored blue, red, and black respectively.

For all graphs, both nodes and edges can be clicked to display details about the chosen target or interaction. The graph may also be resized and moved around for ease of use. A graph query interface has been provided to display the results of both data gathering methods used. In a graph query page, users can select two SDGs to generate a graph for. The system will then display a graph showing all the target interactions between those two SDGs. An example of a graph query view for a network of SDG target interactions can be seen in Appendix: Figure 5.

### 3.2.2. Other Featured Pages

Other featured pages display the current results of the study in text or tabular form for readability. For both data gathering methods, a page listing all target pairs with negative/trade-off interactions is provided as well as a page for positive/synergistic interactions. Another page displays a list of all targets with their descriptions, color-coded to show whether they are beautiful or ugly. As discussed previously, beautiful targets, those that do not have any negative interactions, are colored blue while ugly targets are colored red. Those that do not have an evaluation yet are colored black.

## IV. Results

Since two methods for data gathering have been used, results from both methods will first be discussed separately then synthesized in the next section in order to draw conclusions in the form of a positive and negative answer.

### 4.1. First Method: Expert Evaluation

From a total of 169 SDG targets, the total number of interactions amount to 14196 (169*168/2). Of these interactions represented by graph edges, 1256 (8.85%) edges have been colored so far; of which 36 (2.87%) are negative, 981 (78.11%) are positive, and 239 (19.03%) are zero (consistent).

4.1.1. Intra-goal Target Interactions

4.1.1.1. Negative

Among the evaluated negative interactions, 12 are intra-goal interactions. These negative intra-goal target interactions can be seen in Appendix: List 1. These interactions fall under SDG 3 "Good Health and Well-being", SDG 4 "Quality Education", SDG 5 "Gender Equality", SDG 8 "Decent Work and Economic Growth", SDG 10 "Reduced Inequality", SDG 12 "Responsible Consumption and Production", and SDG 16 "Peace, Justice, and Strong Institutions".

SDG 16 has the most negative intra-goal target interactions, totaling to four. It is followed by SDG 3 which has three while the rest of the involved SDGs have only one each.

[Explanation regarding why SDG 16 has many negative intra-goal interactions.]

4.1.1.2. Positive

Among the evaluated positive interactions, 315 are intra-goal interactions. These positive intra-goal target interactions can be seen in Appendix: List 2. These interactions are distributed across all SDGs, with the top three being SDG 3 "Good Health and Well-being" having 35, SDG 8 "Decent Work and Economic Growth" having 33, and SDG 16 "Peace, Justice, and Strong Institutions" having 29.

[Explanation regarding why SDG 3 has many positive intra-goal interactions.]

4.1.2. Ugly and Beautiful targets

From the current results, 116 of the 169 targets (68.7%) are beautiful, 51 (30.2%) are ugly, and two (1.1%) do not have any evaluated target interactions.

4.1.2.1. Ugliest Targets

Among the ugly targets, 15 have multiple negative interactions. The list of these ugly targets along with the number of their negative interactions can be seen in Appendix: List 3.

Target 13.1 "Strengthen resilience and adaptive capacity to climate related disasters" has the largest number of negative interactions with other targets, totaling to four, making it the ugliest target. It is followed by targets 5.B, 8.2, 12.4, and 16.1 which have three each.

[Explanation regarding why Target 13.1 might have so many negative interactions.]

4.1.2.2. Most Beautiful Targets

Among the beautiful targets, 108 have multiple positive interactions. The list of these beautiful targets along with the number of their positive interactions can be seen in Appendix: List 4.

Target 7.1 "Universal access to modern energy" has the largest number of positive interactions with other targets, totaling to 65, making it the most beautiful target. It is followed by targets 1.3 and 5.5 which have 31 each.

[Explanation regarding why Target 7.1, 1.3, and 5.5 might have so many positive interactions.]

## 4.2. Second Method: Official UN Indicator Data

Among the total 14196 target interactions, 292 (2.06%) are evaluated to be synergies while 236 (1.66%) are trade-offs. The rest of the interactions are considered non-classified either because there is no sufficient data or a majority of the resulting Spearman rank correlation coefficients at the target level fall between -0.6 and 0.6.

### 4.2.1. Intra-goal Target Interactions

#### 4.2.1.1. Trade-Offs

Among the evaluated trade-offs, 23 are intra-goal interactions. These trade-off intra-goal target interactions can be seen in Appendix: List 5. These interactions fall under SDG 1 "No Poverty", SDG 3 "Good Health and Well-being", SDG 7 "Affordable and Clean Energy", SDG 9 "Industry, Innovation, and Infrastructure", SDG 10 "Reduced Inequalities", SDG 15 "Life on Land", and SDG 17 "Partnership for the Goals".

SDG 3 has the most trade-off intra-goal target interactions, totaling to ten. It is followed by SDG 17 which has four then SDGs 7, 9, 10, and 15 which have two each.

[Explanation regarding why SDG 3 has many negative intra-goal interactions.]

#### 4.2.1.2. Synergies

Among the evaluated synergies, 21 are intra-goal interactions. These synergistic intra-goal target interactions can be seen in Appendix: List 6. These interactions fall under SDG 1 "No Poverty", SDG 3 "Good Health and Well-being", SDG 4 "Quality Education", SDG 6 "Clean Water and Sanitation", SDG 7 "Affordable and Clean Energy", SDG 8 "Decent Work and Economic Growth", SDG 9 "Industry, Innovation, and Infrastructure", SDG 15 "Life on Land", and SDG 17 "Partnership for the Goals".

SDG 3 also has the most synergistic intra-goal interactions, totaling to five. It is followed by SDGs 8 which has four then SDGs 15 and 17 which have three each.

[Explanation regarding why SDG 3 has many positive intra-goal interactions. Same as with the previous method.]

### 4.2.2. Ugliest and Most Beautiful targets

From the current results, 110 of the 169 targets (65.09%) are beautiful while 59 (34.91%) are ugly.

#### 4.2.2.1. Ugliest Targets

Among the ugly targets, 54 have multiple negative interactions. The list of all the ugly targets along with the number of their negative interactions can be seen in Appendix: List 7.

Target 3.4 "Reduce mortality from non-communicable diseases and promote mental health" has the largest number of negative interactions with other targets, totalling to 27, making it the ugliest target. It is followed by targets 10.6 "Enhanced representation for developing countries in financial institutions" and 16.8 "Strengthen the participation in global governance" having 26 each.

[Explanation regarding why Targets 3.8, 15.5, and 7.2 might have so many negative interactions.]

#### 4.2.2.2. Most Beautiful Targets

Among the beautiful targets, only two have synergistic interactions with the rest being non-classified. The targets are namely, target 8.5 "Full employment and decent work with equal pay" and target 17.5 "Invest in least developed countries".

[Explanation regarding why targets 8.5 and 17.5 are good to focus on.]

# V. Discussion and Recommendations

### 5.1. Novelty of the Study

What makes this study different from existing studies is that opposed to just using one method, either expert evaluation or official indicator data, this study makes use of both. Results from both are compared and common targets/target pairs are identified in order to have strong recommendation backed by both data gathering methods with regards to what targets/target pairs need to be prioritized and which ones need to be resolved.

### 5.2. Integration of Results

Since available data or responses are limited for both methods, we still have a long way to go to evaluate all 14196 target interactions. More respondents need to be recruited to provide expert evaluations for the first method, and because updates to official indicator data are not conducted wholly, that is only select indicators are updated at a time, data used should be filled in and updated as it becomes available.

The responses and indicator data that is available though is sufficient to formulate an initial recommendation for policymakers in the form of a positive and negative answer tailored to the Philippine setting.

One insight that can be gained from the results so far is that most target interactions are either positive or neutral. This is a good sign that supports the 2030 Agenda in that it is largely applicable to the Philippine context. This also means that there are relatively few negative interactions that need to be resolved.

5.2.1. Negative Answer

It can be observed that SDG 3 "Good Health and Well-being" is the only common goal between the two data-gathering methods that has negative/trade-off intra-goal interactions with three under the first method and ten under the second method. While none of the target pairs are common, we can narrow down as a focus for resolution the commonly involved targets 3.1 "Reduce maternal mortality", 3.6 "Reduce road injuries and deaths", and 3.7 "Universal access to sexual and reproductive care, family planning, and education". Target 3.6 in particular is problematic, having conflict with seven other SDG 3 targets: 3.1, 3.4, 3.5, 3.7, 3.8, 3.9, and 3.D.

Among the ugly targets that have multiple negative interactions, the following targets are common among the two methods: 3.6, 3.7, and 8.2 "Diversify, innovate, and upgrade for economic productivity".

[Explanation for why targets 3.1, 3.6, 3.7, and 8.2 are problematic.]

5.2.2. Positive Answer

In terms of positive/synergistic intra-goal target interactions, the common SDGs across both methods are SDG 1 "No Poverty", SDG 3 "Good Health and Well-being", SDG 4 "Quality Education", SDG 6 "Clean Water and Sanitation", SDG 8 "Decent Work and Economic Growth", SDG 9 "Industry, Innovation, and Infrastructure", the following target pairs are also common:

- 1.1. – 1.2
- 3.7 – 3.1, 3.2
- 4.2 – 4.B
- 6.2 – 6.6
- 8.1 – 8.5
- 9.4 – 9.5

[Explanation for why the above target pairs work well together.]

Furthermore, the common beautiful targets across both methods were: 8.5 "Full employment and decent work with equal pay" and 17.5 "Invest in least developed countries".

**5.3. Scope and Limitations**

Experts who have contributed to the study hail mostly from (1) the University of the Philippines College of Social Work and Community Development and (2) the UP National College of Public Administration and Governance in the University of the Philippines—Diliman. Their responses, in the form of SDG interaction scores dated up until March 2022, were analyzed in this study.

For the official indicator data method, results have been retrieved from a study conducted by Anderson (2021) wherein a Spearman rank correlation was run on publicly available UN indicator data from 2018. Future updates to official indicator data must be used to keep the current analysis up-to-date.

Results of the study are limited to the Philippine setting although the methodology may be adapted to other smaller or larger contexts at risk of difficulty in data gathering.

# VI. Summary and Conclusion

**6.1. Implementation of SDGs**

Since the ultimate goal of the study is to provide a guide towards sustainable development in the Philippines, the results of the study from the initial data can already be considered. Recommendations based on the results translate to problematic targets/target pairs whose interactions need to be resolved and good targets/targets pairs to be prioritized.

6.1.1. Resolution of Negatives

It appears that there should be an effort to resolve negative intra-goal target interactions involving targets 3.1 "Reduce maternal mortality", 3.6 "Reduce road injuries and deaths", and 3.7 "Universal access to sexual and reproductive care, family planning, and education".

Common ugly targets with multiple negative interactions should also be resolved, namely, 3.6, 3.7, and 8.2 "Diversify, innovate, and upgrade for economic productivity".

[Recommendation on how conflicts arising from Targets 3.1, 3.6, 3.7, and 8.2 can be resolved.]

Even though they may not be common across both data gathering methods, efforts should also be made to investigate the ugliest targets such as targets 3.4 "Reduce mortality from non-communicable diseases and promote mental health", 10.6 "Enhanced representation for developing countries in financial institutions", and 16.8 "Strengthen the participation in global governance".

6.1.2. Prioritization of Positives

An insight from the results so far is that most target interactions are either positive/synergistic or zero/nonclassified. This is a good sign that the 2030 Agenda is, for the most part, applicable to the Philippine context. Since target interactions are mostly positive, focus should thus be narrowed down to targets that reinforce their corresponding SDGs by having positive/synergistic interactions. These targets include 1.1 "Eradicate extreme poverty", 1.2 "Reduce poverty by at least 50%", 4.2 "Equal access to quality pre-primary education", 4.B "Expand higher education scholarships for developing countries", 6.2 "End open defecation and provide access to sanitation and hygiene", 6.6 "Protect and restore water-related ecosystems", 8.1 "Sustainable economic growth", 8.5 "Full employment and decent work with equal pay", 9.4 "Upgrade all industries and infrastructures for sustainability", and 9.5 "Enhance research and upgrade industrial technologies".

Note that targets that reinforce target 3.7 are excluded since target 3.7 is already listed as a problematic target in the negative answer.

[Recommendation on how to implement reinforcement of target pairs 1.1-1.2, 4.2-4.B, 6.2-6.6, 8.1-8.5, 9.4-9.5.]

Common beautiful targets across the two data-gathering methods should also be prioritized, namely, 8.5 "Full employment and decent work with equal pay" and 17.5 "Invest in least developed countries". Note that target 8.5 is both synergistic with another target under SDG 8 and is also a beautiful target. Thus, among the priority targets, target 8.5 should be given special interest.

[Recommendation on how to focus efforts on target 8.5.]

Focus should also be given to beautiful targets with an unusually high number of positive interactions from either method such as targets 7.1 "Universal access to modern energy", 1.3 "Implement social protection systems", and 5.5 "Ensure full participation in leadership and decision-making".

## 6.2. Further study on SDG target interactions

Future pursuits of this study should focus on gathering more data for both data gathering methods to have a more complete analysis. Another thing to consider, as noted in [Pradhan et al., 2017] would be the possibility of false trade-offs and synergies arising from very little data or seemingly conflicting targets/indicators that actually reinforce each other and vice versa.

## Acknowledgment

The authors wish to thank Dr. Prajal Pradhan of the Potsdam Institute for Climate Impact Research for sharing his data with us and many fruitful scientific discussions. He gave us a very interesting direction to take in our study of the SDGs.

Valdez, Arian Allenson. SDG Target Interactions: The Philippine Analysis of Indivisible and Cancelling Targets. *arXiv*:2109.05532 (2021).

**Appendix**

Figure 1. User Sign-up Interface

Figure 2. User Add Goals Interface

Figure 3. User Survey Module Interface

Figure 4. The 17 SDGs of Agenda 2030

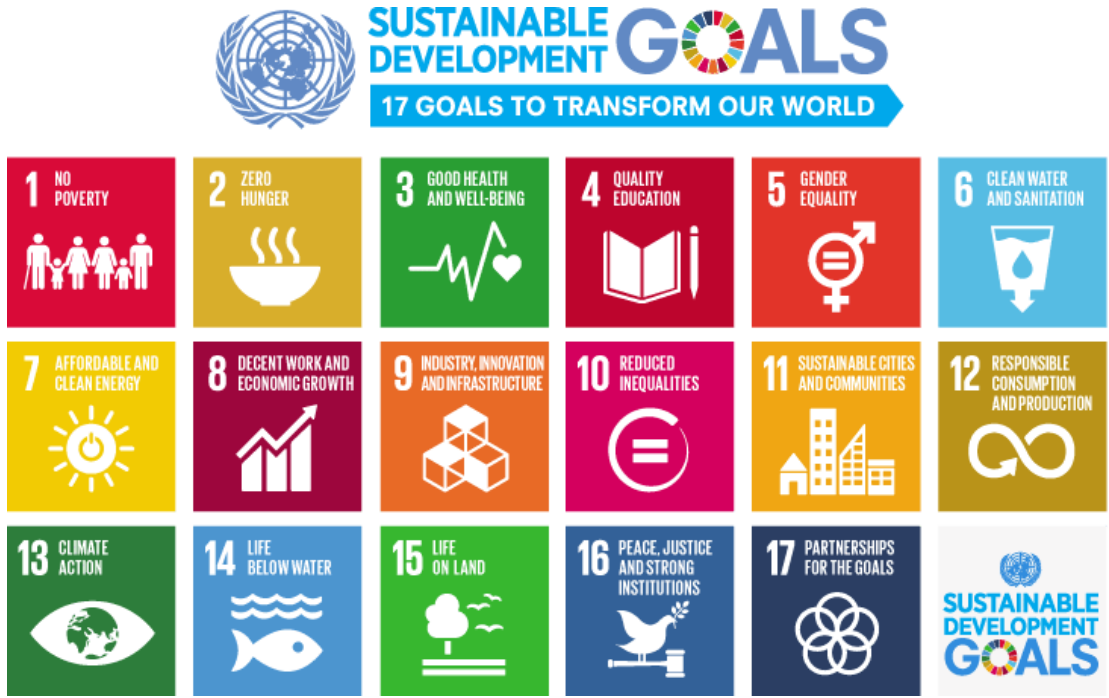

Figure 5. Graph Query Interface

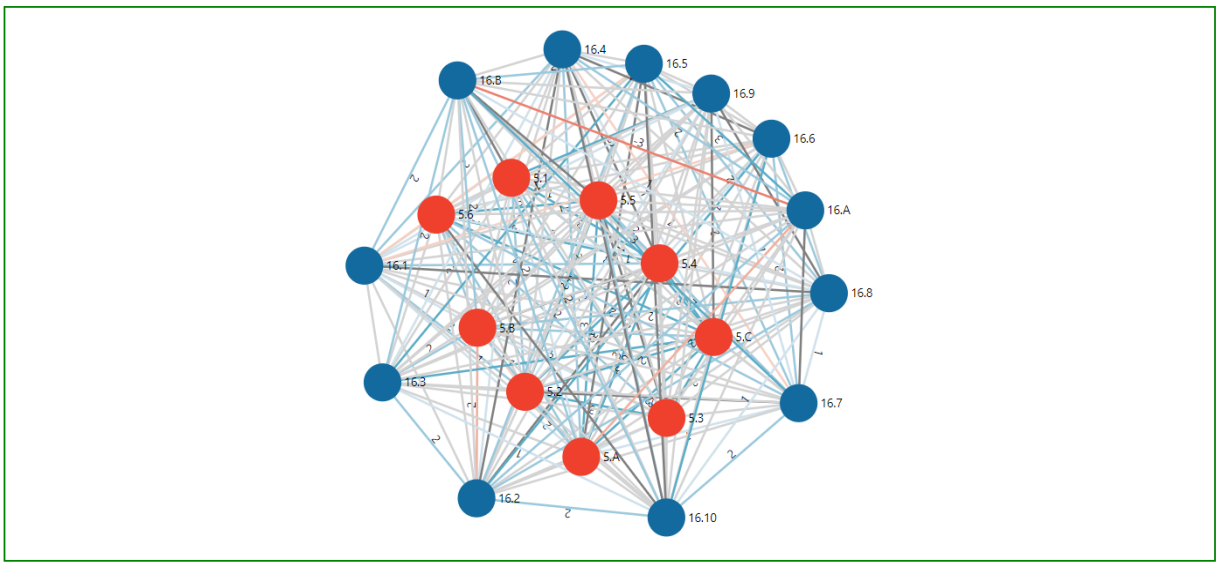

List 1: Negative Intra-Goal Target Interactions under Expert Evaluation Method. *See section 4.1.1.1.*

- SDG 3 (3)
    - 3.6 – 3.1, 3.9
    - 3.7 – 3.9
- SDG 4 (1)
    - 4.7 – 4.B
- SDG 5 (1)
    - 5.2 – 5.B
- SDG 8 (1)
    - 8.2 – 8.6
- SDG 10 (1)
    - 10.5 – 10.C
- SDG 12 (1)
    - 12.1 – 12.4
- SDG 16 (4)
    - 16.1 – 16.5, 16.6
    - 16.4 – 16.7
    - 16.A – 16.B

List 2: Positive Intra-Goal Target Interactions under Expert Evaluation Method. *See section 4.1.1.2.*

- SDG 1 (19)
    - 1.2 – 1.1, 1.3, 1.4, 1.5, 1.A, 1.B
    - 1.A – 1.1, 1.3, 1.4, 1.5, 1.B
    - 1.B – 1.1, 1.3, 1.4, 1.5
    - 1.1 – 1.4, 1.5
    - 1.3 – 1.4, 1.5
- SDG 2 (18)
    - 2.A – 2.1, 2.2, 2.3, 2.5, 2.B, 2.C
    - 2.4 – 2.1, 2.3, 2.5, 2.B, 2.C
    - 2.3 – 2.5, 2.B, 2.C
    - 2.2 – 2.5, 2.B
    - 2.C – 2.1, 2.5
- SDG 3 (35)
    - 3.5 – 3.1, 3.2, 3.3, 3.6, 3.9, 3.A, 3.B, 3.C, 3.D
    - 3.D – 3.1, 3.2, 3.3, 3.8, 3.A, 3.B, 3.C
    - 3.A – 3.4, 3.8, 3.9, 3.B, 3.C
    - 3.7 – 3.1, 3.2, 3.8, 3.B
    - 3.3 – 3.1, 3.8, 3.9
    - 3.6 – 3.2, 3.4, 3.B
    - 3.8 – 3.9, 3.C
    - 3.2 – 3.9
    - 3.4 – 3.B
- SDG 4 (27)
    - 4.2 – 4.1, 4.3, 4.4, 4.6, 4.7, 4.A, 4.B
    - 4.3 – 4.4, 4.5, 4.6, 4.7, 4.B, 4.C

- 4.A – 4.4, 4.6, 4.7, 4.B, 4.C
- 4.C – 4.4, 4.5, 4.6, 4.B
- 4.1 – 4.4, 4.5, 4.B
- 4.4 – 4.6
- 4.5 – 4.7
- SDG 5 (28)
    - 5.A – 5.1, 5.2, 5.3, 5.4, 5.5, 5.6, 5.B, 5.C
    - 5.C – 5.1, 5.2, 5.3, 5.4, 5.5, 5.6, 5.B
    - 5.4 – 5.1, 5.2, 5.3, 5.5, 5.6
    - 5.1 – 5.2, 5.3, 5.5
    - 5.6 – 5.3, 5.5, 5.B
    - 5.2 – 5.5
    - 5.3 – 5.B
- SDG 6 (15)
    - 6.1 – 6.2, 6.3, 6.4, 6.5, 6.6, 6.A
    - 6.6 – 6.2, 6.3, 6.4, 6.A, 6.B
    - 6.4 – 6.2, 6.B
    - 6.5 – 6.3, 6.B
- SDG 7 (2)
    - 7.A – 7.1, 7.3
- SDG 8 (33)
    - 8.5 – 8.1, 8.2, 8.3, 8.4, 8.6, 8.10, 8.A, 8.B
    - 8.9 – 8.1, 8.2, 8.3, 8.4, 8.6, 8.7, 8.8
    - 8.A – 8.1, 8.3, 8.4, 8.6, 8.7, 8.10
    - 8.1 – 8.4, 8.7, 8.8, 8.10
    - 8.3 – 8.7, 8.8, 8.10, 8.B
    - 8.2 – 8.8, 8.B
    - 8.7 – 8.4, 8.6
- SDG 9 (11)
    - 9.1 – 9.2, 9.3, 9.5, 9.C
    - 9.3 – 9.5, 9.A, 9.C
    - 9.2 – 9.A, 9.B
    - 9.4 – 9.5, 9.B
- SDG 10 (17)
    - 10.4 – 10.2, 10.3, 10.5, 10.7, 10.B, 10.C
    - 10.1 – 10.2, 10.5, 10.B
    - 10.3 – 10.6, 10.A, 10.C
    - 10.6 – 10.5, 10.B
    - 10.7 – 10.2, 10.A
    - 10.A – 10.C
- SDG 11 (22)
    - 11.5 – 11.1, 11.2, 11.3, 11.4, 11.6, 11.A, 11.C
    - 11.2 – 11.1, 11.6, 11.7, 11.A, 11.B
    - 11.3 – 11.1, 11.4, 11.7, 11.A
    - 11.1 – 11.6, 11.A

- o 11.7 – 11.6, 11.C
- o 11.4 – 11.A
- o 11.B – 11.C
- SDG 12 (26)
  - o 12.6 – 12.1, 12.2, 12.3, 12.4, 12.8, 12A
  - o 12.5 – 12.3, 12.4, 12.8, 12.B, 12.C
  - o 12.1 – 12.2, 12.7, 12.A, 12.C
  - o 12.8 – 12.4, 12.A, 12.B, 12.C
  - o 12.3 – 12.7, 12.A, 12.C
  - o 12.2 – 12.A, 12.B
  - o 12.C – 12.7, 12.B
- SDG 13 (5)
  - o 13.B – 13.1, 13.3, 13.A
  - o 13.2 – 13.1, 13.3
- SDG 14 (12)
  - o 14.1 – 14.2, 14.3, 14.4, 14.5, 14.A, 14.B
  - o 14.5 – 14.3, 14.7, 14.A, 14.B
  - o 14.6 – 14.B
  - o 14.7 – 14.A
- SDG 15 (2)
  - o 15.6 – 15.1, 15.2
- SDG 16 (29)
  - o 16.2 – 16.3, 16.5, 16.6, 16.8, 16.10, 16.A
  - o 16.7 – 16.5, 16.6, 16.8, 16.9, 16.10, 16.B
  - o 16.B – 16.1, 16.3, 16.5, 16.8, 16.10
  - o 16.4 – 16.1, 16.3, 16.9, 16.A
  - o 16.5 – 16.3, 16.8, 16.A
  - o 16.10 – 16.3, 16.6, 16.8
  - o 16.1 – 16.9
  - o 16.6 – 16.8
- SDG 17 (14)
  - o 17.14 – 17.5, 17.6, 17.7, 17.13, 17.17
  - o 17.2 – 17.1, 17.4
  - o 17.5 – 17.3, 17.10
  - o 17.7 – 17.6, 17.10
  - o 17.12 – 17.11, 17.16
  - o 17.8 – 17.19

List 3: Ugly Targets with Multiple Negative Interactions under Expert Evaluation. *See section 4.1.2.1*.

- 3.6 (2)
- 3.7 (2)
- 3.9 (2)
- 3.A (2)
- 5.A (2)
- 5.B (3)
- 8.2 (3)
- 10.C (2)
- 11.6 (2)

- 12.4 (3)
- 13.1 (4)
- 16.1 (3)
- 16.2 (2)
- 16.7 (2)
- 16.A (2)

List 4: Beautiful Targets with Multiple Positive Interactions under Expert Evaluation Method. *See section 4.1.2.2.*

- 1.1 (26)
- 1.2 (25)
- 1.3 (31)
- 1.4 (22)
- 1.B (27)
- 2.1 (9)
- 2.2 (7)
- 2.3 (15)
- 2.4 (12)
- 2.5 (10)
- 2.A (16)
- 2.B (16)
- 2.C (13)
- 3.2 (10)
- 3.3 (11)
- 3.4 (11)
- 3.5 (15)
- 3.8 (17)
- 3.B (18)
- 3.C (11)
- 3.D (17)
- 4.1 (9)
- 4.2 (13)
- 4.3 (14)
- 4.4 (10)
- 4.5 (20)
- 4.6 (10)
- 4.A (11)
- 4.C (15)
- 5.5 (31)
- 5.6 (26)
- 5.C (26)
- 6.1 (26)
- 6.2 (10)
- 6.4 (11)
- 6.5 (10)
- 6.A (15)
- 7.1 (65)
- 7.2 (7)
- 7.3 (4)
- 7.A (7)
- 7.B (7)
- 8.3 (13)
- 8.5 (17)
- 8.7 (10)
- 8.8 (8)
- 8.9 (16)
- 8.10 (14)
- 8.A (17)
- 8.B (7)
- 9.1 (11)
- 9.2 (7)
- 9.3 (14)
- 9.4 (6)
- 9.5 (13)
- 9.A (6)
- 9.B (8)
- 9.C (11)
- 10.2 (10)
- 10.3 (13)
- 10.4 (19)
- 10.7 (7)
- 10.A (8)
- 10.B (7)
- 11.1 (11)
- 11.2 (20)
- 11.5 (18)
- 11.7 (16)
- 11.A (17)
- 11.B (9)
- 11.C (9)
- 12.6 (12)
- 12.8 (28)
- 12.A (15)
- 12.B (15)
- 12.C (16)
- 13.2 (9)
- 13.A (7)
- 13.B (8)
- 14.1 (7)
- 14.2 (2)
- 14.3 (5)
- 14.4 (5)
- 14.6 (5)
- 14.7 (5)
- 14.A (6)
- 14.B (6)
- 15.2 (2)
- 15.6 (2)
- 15.9 (4)
- 15.B (3)
- 16.3 (12)
- 16.9 (11)
- 16.10 (14)
- 17.1 (4)
- 17.4 (2)
- 17.5 (4)
- 17.6 (5)
- 17.7 (4)
- 17.8 (2)
- 17.9 (3)
- 17.10 (4)
- 17.12 (8)
- 17.13 (5)
- 17.14 (9)
- 17.16 (3)
- 17.17 (8)
- 17.18 (6)
- 17.19 (7)

List 5: Trade-Off Intra-Goal Target Interactions under Official Indicator Data Method. *See section 4.2.1.1*.

- SDG 1
  - 1.1 – 1.A
- SDG 3
  - 3.4 – 3.1, 3.2, 3.6, 3.D
  - 3.6 – 3.5, 3.7, 3.8, 3.D

- - 3.2 – 3.5, 3.8
- SDG 7
  - 7.2 – 7.1, 7.3
- SDG 9
  - 9.3 – 9.C
  - 9.5 – 9.A
- SDG 10
  - 10.6 – 10.4, 10.B
- SDG 15
  - 15.5 – 15.1, 15.4
- SDG 17
  - 17.6 – 17.3, 17.4
  - 17.4 – 17.8, 17.9

List 6: Synergistic Intra-Goal Target Interactions under Official Indicator Data Method. *See section 4.2.1.2.*

- SDG 1
  - 1.1 – 1.2
- SDG 3
  - 3.1 – 3.7, 3.C
  - 3.2 – 3.3, 3.7
  - 3.4 – 3.5
- SDG 4
  - 4.2 – 4.B
- SDG 6
  - 6.2 – 6.6
- SDG 7
  - 7.1 – 7.3
- SDG 8
  - 8.1 – 8.2, 8.5
  - 8.10 – 8.2, 8.6
- SDG 9
  - 9.5 – 9.4, 9.B
- SDG 15
  - 15.1 – 15.A, 15.B
  - 15.A – 15.B
- SDG 17
  - 17.6 – 17.8, 17.9
  - 17.8 – 17.9

List 7: Ugly Targets under Official Indicator Data Method. *See section 4.2.2.1.*

- 1.1. (9)
- 1.2. (1)
- 1.A. (3)
- 2.1. (5)
- 2.5. (1)
- 2.A. (2)
- 3.1. (7)
- 3.2. (9)
- 3.3. (6)
- 3.4. (27)
- 3.5. (15)
- 3.6. (12)
- 3.7. (5)
- 3.8. (17)
- 3.C. (2)
- 3.D. (4)
- 4.2. (3)
- 4.B. (5)
- 5.5. (1)
- 6.2. (10)
- 6.6. (11)
- 6.A. (6)
- 7.1. (9)
- 7.2. (24)
- 7.3. (11)
- 8.1. (6)
- 8.2. (4)
- 8.4. (6)
- 8.6. (5)
- 8.8. (2)
- 8.10. (7)
- 8.A. (3)
- 9.2. (14)
- 9.3. (8)
- 9.4. (6)
- 9.5. (6)
- 9.A. (5)
- 9.B. (6)
- 9.C. (6)
- 10.4. (4)
- 10.6. (26)
- 10.B. (7)
- 11.1. (9)
- 12.2. (5)
- 14.5. (6)
- 15.1. (4)
- 15.2. (4)
- 15.4. (7)
- 15.5. (25)
- 15.A. (3)
- 15.B. (3)
- 16.8. (26)
- 17.2. (1)
- 17.3. (2)
- 17.4. (19)
- 17.6. (11)
- 17.8. (10)
- 17.9. (10)
- 17.19. (1)